\documentclass[%
 preprint,
 amsmath,amssymb,
 aps,
notitlepage
]{revtex4-2}

\usepackage{graphicx}
\usepackage{dcolumn}
\usepackage{bm}
\usepackage[mathlines]{lineno}

\begin{document}

\preprint{APS/123-QED}

\title{Coherence and superradiance from a plasma-based quasiparticle accelerator}

\author{B. Malaca$^{1}$, M. Pardal$^{1}$, D. Ramsey$^{2}$, J. Pierce$^{3}$, K. Weichman$^{2}$, I. Andriyash$^{4}$, W. B. Mori$^{3}$, J. P. Palastro$^{2}$, R. A. Fonseca$^{1,5}$, J. Vieira$^{1}$}
 \affiliation{$^1$GoLP/IPFN, Instituto Superior Técnico, Universidade de Lisboa, 1049-001 Lisbon, Portugal}
 \affiliation{$^2$University of Rochester, Laboratory for Laser Energetics, Rochester, New York, 14623, USA}
 \affiliation{$^3$Department of Physics and Astronomy, University of California, Los Angeles, CA 90095, USA}
 \affiliation{$^4$LOA, École Polytechnique, ENSTA Paris, CNRS, Institut Polytechnique de Paris, 91762 Palaiseau, France}
 \affiliation{$^5$ISCTE - Lisbon University Institute}

 \begin{abstract}

Coherent light sources, such as free electron lasers, provide bright beams for biology, chemistry, physics, and advanced technological applications. Increasing the brightness of these sources requires progressively larger devices, with the largest being several km long (e.g., LCLS). Can we reverse this trend, and bring these sources to the many thousands of labs spanning universities, hospitals, and industry? Here we address this long-standing question by rethinking basic principles of radiation physics. At the core of our work is the introduction of quasi-particle-based light sources that rely on the collective and macroscopic motion of an ensemble of light-emitting charges to evolve and radiate in ways that would be unphysical when considering single charges. The underlying concept allows for temporal coherence and superradiance in fundamentally new configurations, providing radiation with clear experimental signatures and revolutionary properties. The underlying concept is illustrated with plasma accelerators but extends well beyond this case, such as to nonlinear optical configurations. The simplicity of the quasi-particle approach makes it suitable for experimental demonstrations at existing laser and accelerator facilities.

\end{abstract}

\maketitle

\date{\today}

\newpage

Temporal coherence and superradiance are at the core of the most advanced light sources available today, most notably free electron lasers (FELs)~\cite{bib:gover_prab_2005}. Because the radiation intensity scales favourably with the number of light-emitting particles \textit{squared}, superradiance underlies the success of all the scientific and technological applications enabled by free electron lasers~\cite{bib:shea_science_2001}.

Compact plasma accelerators~\cite{bib:tajima_prl_1979,bib:chen_prl_1985} are a complementary and attractive source of radiation~\cite{bib:rousse_prl_2004}. They provide intrinsically ultra-short, spatially collimated and bright x-rays~\cite{bib:kneip_nphys_2010} for applications in biology~\cite{bib:cole_srep_2015}, high energy density~\cite{bib:wood_srep_2018} and material~\cite{bib:he_srep_2016} science, and nonlinear quantum electrodynamics~\cite{bib:cole_prx_2018}. In contrast to free electron lasers, these sources are compact, not longer than a few cm, but have only produced temporally incoherent radiation, where the intensity scales \textit{linearly} with the number of emitters~\cite{bib:albert_ppcf_2014}.

To reach the brightness of free electron lasers, plasma accelerator-based light sources need to become temporally coherent and superradiant. Such a revolutionary advance could bring research and technology that is only available in a handful of FELs worldwide directly to the many university, hospital, and industrial scale laboratories. Hence, the onset of temporal coherence and superradiance is the essential missing ingredient to make compact, affordable, and competitive plasma accelerator-based light sources.


Recently, free-electron lasing of plasma-accelerated electrons has been demonstrated using conventional magnetic undulators ~\cite{bib:wang_nature_2021, bib:labat_nphotonics_2022, bib:pompilli_nature_2022}. This was enabled by improvements in plasma-based acceleration that led to the generation of GeV-class electron bunches with sufficiently low emittances and energy spreads~\cite{bib:wang_prl_2006}. These are remarkable achievements, but a lasing process that 
relies only on the plasma itself is essential in order to miniaturize these light sources.

\begin{figure}[h!]
\includegraphics[width=1.0\columnwidth]{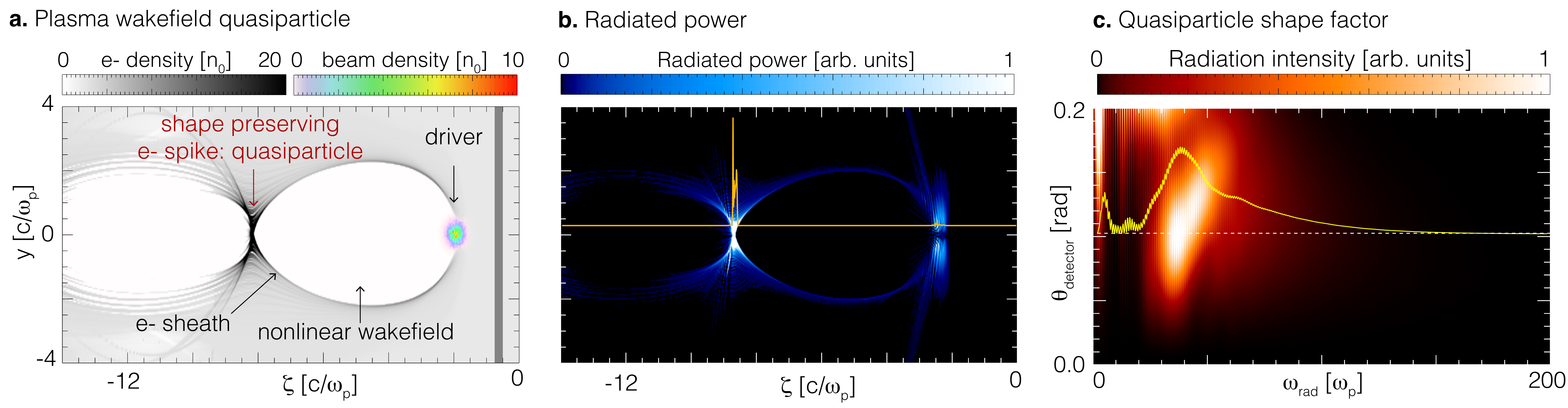}
\caption{\label{fig:quasiparticle} Plasma wakefield quasiparticles and their radiation. a. Three-dimensional particle-in-cell simulation using Osiris (see methods) where an electron beam drives a nonlinear plasma wakefield in the blowout regime. The plasma is in gray, and the electron bunch driver is in rainbow colours. The electron spike at the back of the plasma oscillation is a nearly propagation-invariant structure that plays the role of a light-emitting quasiparticle. b. Radiated power from the nonlinear wakefield in a. An horizontal lineout is in orange. Colours are saturated to improve visualisation. c. shows the shape function $|\mathbf{S}(\omega)|^2$ for the nonlinear wakefield quasiparticle, determined using the post-processing radiation algorithm RaDiO (see methods). This corresponds to the radiation emitted by a thin electron slice in $\xi$ as it crosses the electron bunch driver and returns to the axis. The radiation was recorded in a virtual detector placed in the far field. The yellow line shows a lineout of $|\bm{\mathcal{S}}(\omega)|^2$ at $\theta = 0.1~\mathrm{rad}$.
}
\end{figure}

Lasing processes based purely on plasma-accelerators, without the need of a conventional undulator, require  electron bunches with even higher quality than those used in an FEL ~\cite{bib:davoine_jpp_2018}. Such bunches, however, are not yet available. Here, we circumvent this challenge by introducing a light source concept based on the motion of quasiparticle excitations, which depends on the coordinated motion of an ensemble of light-emitting particles~\cite{bib:malaca_aps_2022,bib:malaca_eps_2022, bib:malaca_aac_2022, bib:malaca_hedla_2022}. This concept introduces a new paradigm in which temporal coherence and superradiance do not directly depend on the bunch quality. Instead, the temporal coherence and superradiance require a localized current density with a near-constant profile, such as those routinely generated in the wake of intense lasers and particle bunches in plasma. Figure~\ref{fig:quasiparticle} demonstrates the quasiparticle concept. We denote such localized, near-constant current density profiles as quasiparticles.


Because they result from collective motion, quasiparticles can propagate at any velocity $v_c$, including superluminal ($v_c>c$), and can be subject to any acceleration. We show that experimentally demonstrated plasma density structures ~\cite{bib:burza_prab_2013,bib:geddes_prl_2008,bib:gonsalves_nphys_2011,bib:layer_prl_2007,bib:layer_oe_2009} can be used to accurately control the quasiparticle trajectory. The flexibility to control the quasiparticle trajectory enables fundamentally new radiation physics that features the essential ingredients of temporal coherence and superradiance. Examples include superradiant Cherenkov emission from superluminal quasiparticles and a novel superluminal undulator radiation regime, which is classically forbidden for point-like charges. These examples can be achieved using plasma accelerators, but the underlying concept can be extended to excitations based on bound charges, as in nonlinear optics, or through perturbations to the magnetization. Here, we focus on a set of examples that can bring temporal coherence and superradiance to plasma accelerator laboratories today and demonstrate that the number of photons produced can provide a clear experimental signature.





To begin our exploration, we consider the intensity radiated by a given current density $\mathbf{j}[\mathbf{r},t]$ per solid angle per unit frequency in the far-field~\cite{bib:jackson}:
\begin{equation}
    \label{eq:radiation}
    \frac{d^2I }{d\omega d\Omega} = (\simeq) \frac{\omega^2}{4\pi^2c^3}~ \left|\int d\mathbf{r}\int dt~\mathbf{n}\times \left[\mathbf{n}\times \mathbf{j}(\mathbf{r},t)\right]~\exp\left[i\omega(t-\mathbf{n}\cdot \mathbf{r}/c)\right]\right|^2,
\end{equation}
where $t$ is the time of emission (retarded time), $\mathbf{r}$ is the position, $\omega$ is the radiation frequency, $\Omega$ the solid angle and $\mathbf{n}$ is a unit vector that sets the observation direction. The observation direction is given, in spherical coordinates, by $\mathbf{n}=[\cos(\theta), \sin(\theta)\sin(\varphi),\sin(\theta)\cos(\varphi)]$, where $\theta$ is the angle with respect to the $x$-axis and $\varphi$ is the angle with respect to the $z$-axis in the $y-z$ plane. 





Equation (\ref{eq:radiation}) applies to arbitrary current density profiles, but here we focus on scenarios where $\mathbf{j}(\mathbf{r},t) = \mathbf{j}[\mathbf{r}-\mathbf{r_c}(t)]$. The expression $\mathbf{r}_c(t)$ could represent the trajectory of a point-like charge, in which case $\mathbf{j}(\mathbf{r},t) \propto \bm{\delta}[\mathbf{r}-\mathbf{r_c}(t)]$. In this work, however, we are interested in the radiation produced by an ensemble of light-emitting particles. Specifically, $\mathbf{j}[\mathbf{r}-\mathbf{r_c}(t)]$ corresponds to a spatially localized current density profile that maintains a constant shape as it moves along the trajectory given by $\mathbf{r}_c(t)$. We view such a propagation-invariant current profile as a finite sized `particle' (i.e., a quasiparticle) which executes the trajectory given by $\mathbf{r}_c(t)$. 

Figure~\ref{fig:quasiparticle}a demonstrates the concept of a quasiparticle in the context of plasma-based accelerators. In Fig.~\ref{fig:quasiparticle}a, an ultra-relativistic electron bunch radially expels all of the plasma electrons from its path. Most of the radially expelled electrons accumulate in a thin sheath which crosses the axis periodically, leading to a strongly nonlinear wakefield in the so-called blowout regime~\cite{bib:pukhov_apb_2002,bib:lu_prl_2007}. When the sheath electrons cross the axis, they form sub-plasma-skin-depth density spikes that produce most of the radiation in the wakefield in the absence of trapping (see Fig.~\ref{fig:quasiparticle}b). These spikes are characterised by a current density that maintains a near-constant shape and can therefore be regarded as quasiparticles.

To illustrate the key features of quasiparticle radiation, we change to the co-moving frame, where [$\bm{\xi} = \mathbf{r}-\mathbf{r_c}(\tau), \tau=t$]. In this new set of coordinates $(\bm{\xi},\tau)$ that move with $\mathbf{r}_c(t)$, Eq.~(\ref{eq:radiation}) can be re-written in a more suggestive form as:
\par
\begin{equation}
    \label{eq:collective}
    \frac{d^2I }{d\omega d\Omega}=\frac{\omega^2}{4\pi^2c^3}~ \left|\int d\tau \bm{\mathcal{S}}(\omega) \exp [i\omega(\tau-\mathbf{n}\cdot\mathbf{r_c}(\tau)/c)]\right|^2,
\end{equation}
where $\bm{\mathcal{S}}(\omega)$ represents a shape factor given by: 
\begin{equation}
    \label{eq:weight}
    \bm{\mathcal{S}}(\omega) =  \int d\bm{\xi}~\mathbf{n}\times \left[\mathbf{n}\times \mathbf{j}(\bm{\xi})\right] \exp [-i\omega~\mathbf{n}\cdot \bm{\xi}/c].
\end{equation}
When combined with Eq.~(\ref{eq:weight}), Eq.~(\ref{eq:collective}) predicts the spectral radiation intensity produced by a quasiparticle. The form of Eq.~(\ref{eq:collective}) is analogous to that of a single point-like charge, except for the shape factor $\bm{\mathcal{S}}(\omega)$, which can be interpreted as the quasiparticle internal spectrum and depends on the specific spatial profile of $\mathbf{j}(\bm{\xi})$. To understand the role of the quasiparticle spatial distribution, we consider first a point-like charge, which is represented in space by a Dirac delta function. In this limit, Eqs.~(\ref{eq:collective}) and (\ref{eq:weight}) recover the single particle result, and $\bm{\mathcal{S}}(\omega)$ is frequency independent. In contrast, the spectrum $\bm{\mathcal{S}}(\omega)$ of a finite-sized quasiparticle has a non-zero but finite bandwidth, as it necessarily vanishes for $\omega\rightarrow \infty$. This is shown in Fig.~\ref{fig:quasiparticle}c for the case of a nonlinear plasma wakefield quasiparticle. The quasiparticle spectrum is characterised by a finite frequency bandwidth, set by the wakefield amplitude, which in Fig.~\ref{fig:quasiparticle} extends up to $\omega_{\mathrm{max}}\simeq 200~\omega_p$. For a plasma with an electron density $n_0=10^{16}-10^{19}~\mathrm{cm}^{-3}$, this gives a range of frequencies from the THz to the extreme ultraviolet.

A key prediction of Eq.~(\ref{eq:collective}) is that the quasiparticle shape and trajectory fully define its radiation spectrum. Surprisingly, this means that quasiparticle radiation is independent of the radiation emitted by its microscopic constituents. To understand why, we focus on the complex phase factor in Eq.~(\ref{eq:collective}), which is exactly the same for a point-like particle and for a quasiparticle. This demonstrates that it is the quasiparticle trajectory that controls interference (just as if it were a single point-like charge), not the trajectories of its individual constituents. 

While the radiation from a quasiparticle and from a point-like charge interfere in the same way, the underlying physics is different: the motion of a quasiparticle is not bound by the same constraints as a classical point particle.  For example, just as electromagnetic energy can flow opposite to its group velocity~\cite{bib:ramsey_pra_2020}, a quasiparticle can travel in the opposite direction as each of its microscopic constituents. This is possible because quasiparticles arise from the collective motion of light-emitting particles, in which individual emitters display coordinated motions relative to each other. As a direct result, quasiparticles can travel at an arbitrary velocity (including superluminal) and be subject to any acceleration (that can be as extreme as that in the vicinity of a black hole). This shifts the focus from particle accelerators, which depend on electric and magnetic fields to control particle trajectories, to a quasiparticle accelerator that relies instead on a collective re-organisation of light-emitting particles.


The flexibility to control the quasiparticle trajectory can bring temporal coherence and superradiance to a multitude of configurations where they would otherwise be impossible. Furthermore, radiating quasiparticles not only reproduce key features of the radiation emitted by a single point-like charge, but they also enable fundamentally new radiation mechanisms. For example, a superluminal quasiparticle can emit Cherenkov radiation in a plasma, which is impossible for a point-like charge because the phase velocity is superluminal. This quasiparticle Cherenkov radiation exhibits a superradiant and single-cycle optical shock directed along the Cherenkov cone-angle. Even more intriguing is the radiation from a superluminally oscillating quasiparticle, which introduces a previously unimagined undulator regime. We now explore both of these examples.

We first focus on the quasiparticle-equivalent of Cherenkov radiation. Consider a quasiparticle traveling with a constant velocity along the longitudinal direction $x$  according to $\mathbf{r}_c(\tau) = v_c \tau \mathbf{e}_x$. Substitution in to Eq.~(\ref{eq:collective}) and integration over $\tau=[-T/2,T/2]$, where $T$ is the total emission time, leads to:
\begin{equation}
    \label{eq:constantv}
    \frac{\mathrm{d}^2I }{\mathrm{d}\omega \mathrm{d}\Omega} =  
     \frac{\omega^2}{4\pi^2c^3} |\bm{\mathcal{S}}(\omega)|^2 T^2 \textrm{sinc}^2\left[\frac{\omega T}{2}\left(1-\frac{v_c \cos\theta}{c}\right)\right],
\end{equation}
where $\mathrm{sinc}(\alpha)=\sin(\alpha)/\alpha$ is a resonance function, and $\alpha = [1-(v_c/c) \cos\theta] (\omega T/2)$ is the corresponding detuning parameter. The resonance function, and thus the radiated intensity, reach their maximum value when $\alpha=0$, or equivalently when:
\begin{equation}
    \label{eq:cherenkov}
1-\frac{v_c\cos\theta }{c}=0\Leftrightarrow \cos\theta \equiv \cos\theta_c = \frac{c}{v_c}.
\end{equation}
Equation~(\ref{eq:cherenkov}) coincides exactly with the well known Cherenkov radiation condition in vacuum for a point-like particle, but it now applies to quasiparticles. Even though their microscopic constituents (e.g., electrons) necessarily propagate with $v<c$, quasiparticles can travel at any speed. It is this peculiar property that enables quasiparticle Cherenkov emission, provided $v_c>c$, whereas point-like particles can never satisfy Eq.~(\ref{eq:cherenkov}). As a result, superluminal quasiparticles can naturally generate an optical shock directed at the Cherenkov cone-angle $\theta_c = \arccos(c/v_c)$. 

The optical shock forms because the phases of multiple light rays, emitted at different times (or, equivalently, by different particles), constructively interfere at the Cherenkov angle. Hence, because radiated fields emitted at different times all interfere constructively at $\theta=\theta_c$ when $\alpha = 0$, the radiated peak intensity is as high as it can possibly be and exhibits a favourable scaling with $T^2$. The peak intensity of this optical shock also scales with the square of the number of radiating particles $N^2$. This is a key property of superradiance. To clarify why, note that $T$ relates to the number of emitters as $N \propto n_0 v_c T \Leftrightarrow T \propto N / (n_0 v_c) \propto N$, with $n_0$ being the number density of light-emitting particles. Hence, the spectral density of the emitted radiation exhibits the typical superradiant scaling given by $\mathrm{d}^2I / (\mathrm{d}\omega \mathrm{d}\Omega) \propto T^2 \propto N^2$. 

\begin{figure}[h!]
\includegraphics[width=1.0\columnwidth]{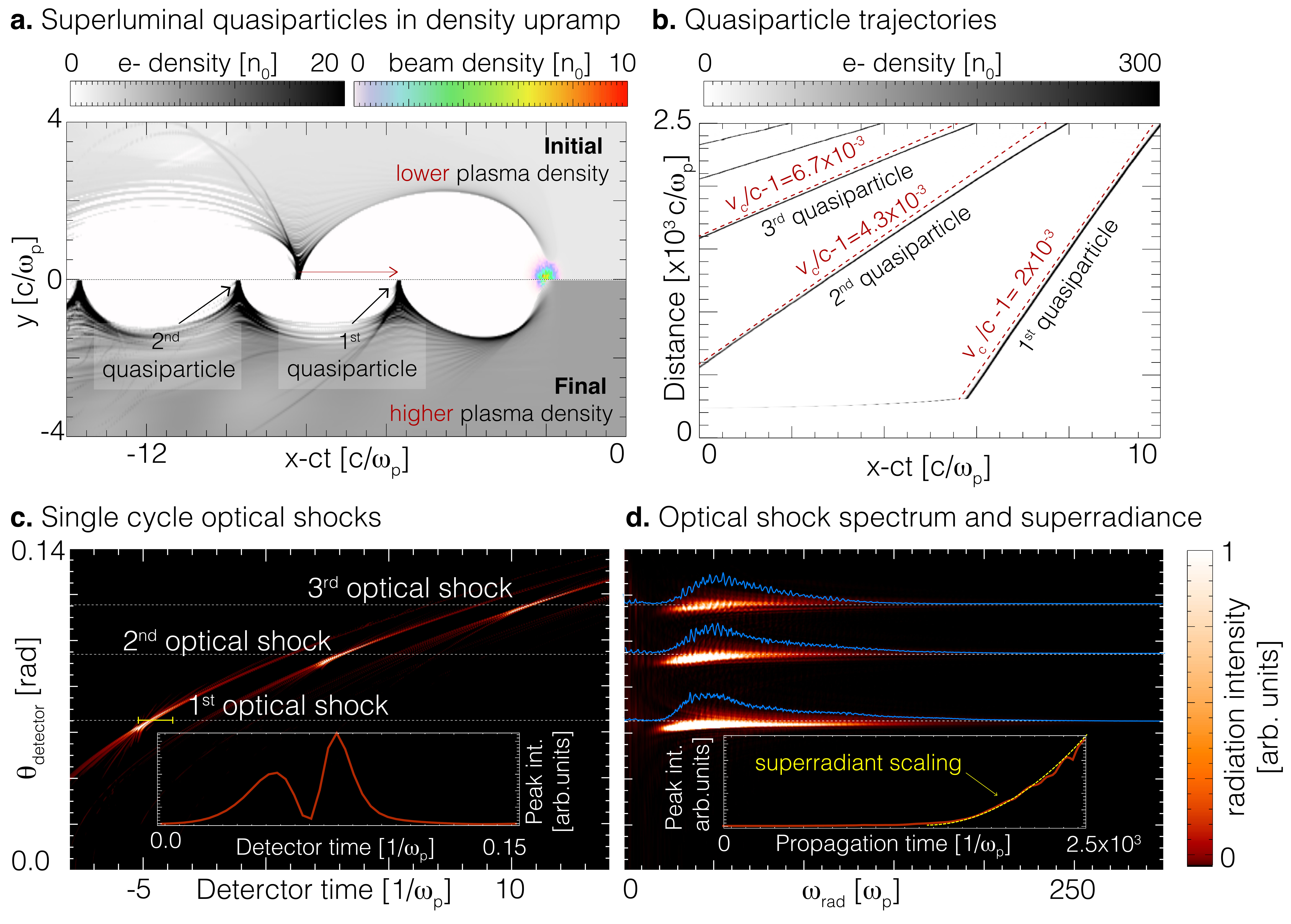}
\caption{\label{fig:cherenkov} Onset of quasiparticle Cherenkov emission. a. Three-dimensional simulation result (see methods) showing plasma wakefields driven by an ultra-relativistic electron bunch in a plasma density upramp. Plasma electron density is in gray, and electron bunch density appears in rainbow colours. Top/bottom frames show the nonlinear wakefield structure earlier/later in the propagation. The quasiparticle trajectory is superluminal because the plasma wavelength decreases for higher plasma density. b. Waterfall plot showing the quasiparticle trajectories in frame moving at $c$. Each horizontal line corresponds to an on-axis lineout of the plasma density as a function of propagation distance. c. Cherenkov radiation optical shocks emitted by three electron spike quasiparticles in a virtual detector placed in the far-field. Colours show radiated intensity, determined numerically using a post-processing radiation tool (see methods). The inset is a closeup that the optical shock consists of a single cycle pulse. d. Corresponding spectral intensity. The inset demonstrates the favourable scaling (yellow) of the peak radiation intensity (red) with propagation distance.
}
\end{figure}

Figure~\ref{fig:cherenkov} illustrates the onset of superradiant quasiparticle Cherenkov emission in a plasma wakefield accelerator. Obtaining precise control over $v_c$ is an essential requirement. Figures~\ref{fig:cherenkov}a-b demonstrate that a spatially varying plasma density profile can control $v_c$ with high accuracy. This control is possible because the local plasma density sets the local plasma wavelength and, as a result, the distance from the driver to the first electron spike. Similar mechanisms, which are based on the so-called accordion effect, were already exploited to induce electron trapping and acceleration in plasma based accelerators~\cite{bib:burza_prab_2013,bib:geddes_prl_2008,bib:gonsalves_nphys_2011,bib:tooley_prl_2017}. It is possible to show that a tailored plasma density profile given by~\cite{bib:xu_prab_2017}:
\begin{equation}
\label{eq:density}
\frac{n(x)}{n_0} = \frac{\lambda_{p0}^2}{[(1-v_c/c) x+ \lambda_{p0}]^2}\simeq \left[1+\frac{2(v_c/c-1)}{\lambda_{p0}} x\right], (1-v_c/c) (x/\lambda_{p0})\ll 1, 
\end{equation}
provides constant quasiparticle velocity over arbitrary propagation distances. This is clearly shown in Fig.~\ref{fig:cherenkov}b, which demonstrates constant-velocity quasiparticle trajectories. In Eq.~(\ref{eq:density}), $n_0$ is the background plasma density at $x=0$, $\lambda_{p0}$ is plasma wavelength at $x=0$, and $x$ is the propagation distance. The derivation of Eq.~(\ref{eq:density}) assumes that the driver travels at $c$ (which is a good approximation for a beam-driven plasma wakefield), and holds for the first electron spike after the driver. The $n^{\mathrm{th}}$ spike after the driver moves approximately with a velocity $v_{c,n}$ given by $(v_c/c-1)=n(v_{c,n}/c-1)$. The quasiparticle velocities, which can be readily deduced from the slopes of the quasiparticle trajectories in Fig.~\ref{fig:cherenkov}b, are in excellent agreement with theoretical predictions. When $(v_c/c-1)(x/m \lambda_p)\ll 1$, $n(x)$ can be Taylor expanded leading to a simple linear plasma density ramp.


Figure~\ref{fig:cherenkov}c shows the far-field radiation intensity profile produced by superluminal quasiparticles. The spatiotemporal intensity profile indicates the presence of three optical shocks, which appear as bright radiation bursts that are both angularly and temporally separated. The optical shocks form at the Cherenkov angle set by the velocity of each quasiparticle, exactly as predicted by Eq.~(\ref{eq:cherenkov}). The inset in Fig.~\ref{fig:cherenkov}c is a magnification of the temporal profile of the first optical shock, which clearly shows that it is a single-cycle optical pulse. The corresponding frequency spectra, shown in Fig.~\ref{fig:cherenkov}d, extend all the way up to $\omega \gtrsim 200~\omega_p$. The inset of Fig.~\ref{fig:cherenkov}d, which shows the peak radiation intensity with propagation distance, illustrates the favourable quadratic scaling typical of superradiance. Hence, this mechanism can produce a train of superradiant, single-cycle, angularly and temporally isolated optical shocks. 

Accelerating quasiparticles can lead to temporal coherence in fundamentally new regimes with no single-particle counterpart. An interesting case, which cannot be realized with point-like charges, is the radiation from a superluminally oscillating quasiparticle. In order to investigate this radiation in a configuration directly accessible to experiments, we consider a purely one-dimensional quasiparticle trajectory. We then assume $\mathbf{r}_c(\tau) = [v_c \tau + \Delta x_c \sin(\omega_c \tau)]\mathbf{e}_x$, where $\Delta x_c$ and $\omega_c$ are the amplitude and frequency of oscillation, respectively. This purely one-dimensional motion contrasts with magnetic undulators, where electrons oscillate in the perpendicular direction. Substituting the expression for $\mathbf{r}_c(\tau)$ into Eq.~(\ref{eq:collective}) yields:
\begin{equation}
\label{eq:undulating}
\frac{\mathrm{d}^2I }{\mathrm{d}\omega \mathrm{d}\Omega} =  \frac{\omega^2}{4\pi^2c^3} T^2 \Bigg|\bm{\mathcal{S}}(\omega)  \sum_{m=-\infty}^{\infty} J_m\left(\frac{\omega \cos(\theta) \Delta x}{c}\right) \textrm{sinc}\left\{\frac{T}{2}\left[\omega\left(1-\frac{v_c\cos\theta}{c}\right)- m \omega_c\right] \right\}\Bigg|^2,
\end{equation} 
where $\mathbf{J}_m(\cdot)$ is the Bessel function of the first kind. The radiation intensity peaks at the resonant frequencies, $\omega_m$, which can be found by setting the argument of the sinc function to zero. This leads to the following expression for $\omega_m$:  
\begin{equation}
\label{eq:resonant}
\omega_m = \frac{m \omega_c}{1-v_c\cos(\theta)}.
\end{equation}
Except for the shape factor $\mathbf{S}(\omega)$, Eqs.~(\ref{eq:undulating}) and Eq.~(\ref{eq:resonant}) coincide exactly with the spectral intensity emitted by a point-like particle propagating along the quasiparticle trajectory $\mathbf{r}_c(\tau)=[v_c \tau + \Delta x_c \sin(\omega_c \tau)]\mathbf{e}_x$. The most crucial, and fundamental, distinction between both cases is that the quasiparticle mean velocity $v_c$ can be arbitrary. This feature leads to a previously unexplored regime of superluminal undulator radiation. 

\begin{figure}[h!]
\includegraphics[width=1.0\columnwidth]{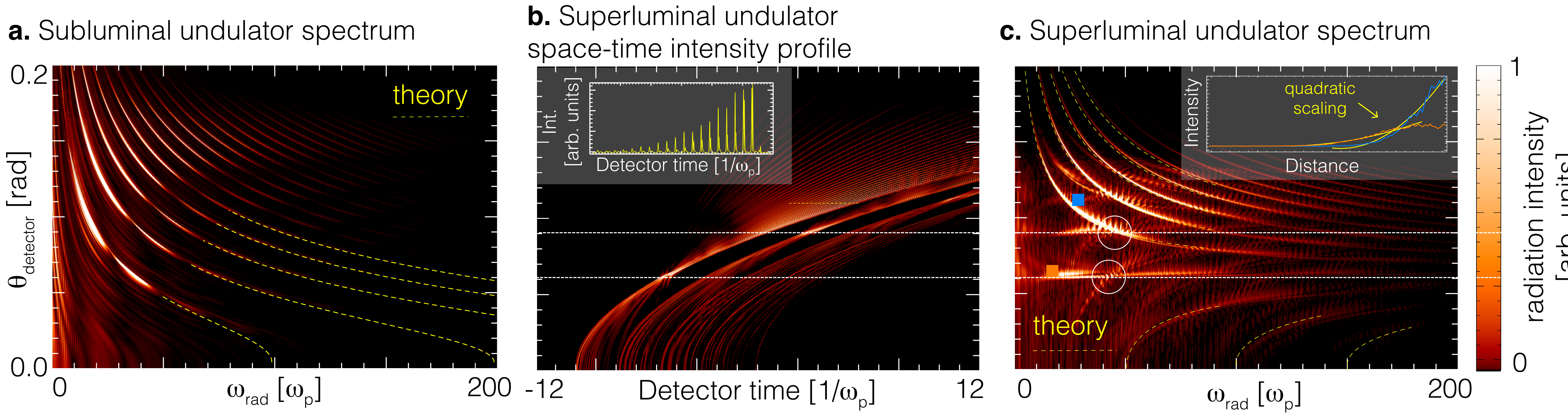}
\caption{\label{fig:undulator} Quasiparticle undulator radiation for an oscillating quasiparticle with $\omega_c = 0.1~\omega_p$ and $\Delta x = 0.25 c/\omega_p$. a. Spectral intensity of undulator radiation from a subluminal quasiparticle with $v_c/c=0.999$. Theoretically calculated resonant frequencies are in yellow. b. Spatiotemporal profile of the radiated intensity in a virtual detector in the far field from superluminal oscillating quasiparticle with $v_c/c=1.0003$. The white dashed line marks the Cherenkov angle corresponding to $v_c$. The inset shows a lineout taken at the interval defined by the yellow dashed line. c. Corresponding spectral radiation intensity. The yellow dashed lines are the theoretically calculated resonant frequencies. The inset shows the evolution of the peak intensity with propagation distance at the positions marked by the squares (Cherenkov angle in orange, first resonant harmonic in blue). The circles show crossing positions between resonant frequency harmonics produced by different quasiparticles.
}
\end{figure}

To investigate this new physics, Fig.~\ref{fig:undulator} demonstrates undulator radiation from oscillating quasiparticles in a nonlinear plasma wakefield driven by an ultra-relativistic electron bunch. By exploiting the so-called accordion effect, the simulations leading to Fig.~\ref{fig:undulator} show that a longitudinally corrugated plasma channel can control the quasiparticle oscillation amplitude and frequency. Furthermore, a plasma density ramp combined with the corrugated plasma channel provides an additional degree of freedom to adjust the mean quasiparticle velocity.

We first investigate undulator radiation from a subluminal quasiparticle, shown in Fig.~\ref{fig:undulator}a. Just as if it were produced by a single charge, the quasiparticle undulator spectrum is more pronounced in the vicinity of the resonant frequencies, which are exactly as predicted by Eq.~(\ref{eq:resonant}). More interstingly, Fig.~\ref{fig:undulator}b-c, which depict the spatiotemporal profile of the radiated intensity (Fig.~\ref{fig:undulator}b) and its corresponding spectral intensity (Fig.~\ref{fig:undulator}c), show a completely new superluminal undulator radiation regime, which is classically forbidden. 

Undulator radiation from superluminal quasiparticles combines features of Cherenkov emission (i.e., an optical shock appears at the angle corresponding to the mean quasiparticle velocity) with undulator radiation harmonics. An optical shock forms due to constructive interference at the Cherenkov angle at $\theta=\mathrm{acos}(1/v_c) \simeq 63~\mathrm{mrads}$ for the first quasiparticle and at $\theta\simeq 90~\mathrm{mrad}$ for the second quasiparticle (second electron spike) in Fig.~\ref{fig:undulator}b. The spectral intensity at the Cherenkov angle is given by the leading order $m=0$ term in Eq.~(\ref{eq:undulating}), leading to $I \propto |\bm{\mathcal{S}}(\omega)|^2 T^2 J_0^2(\omega \Delta x/v_c)$. The inset in Fig.~\ref{fig:undulator}c (orange curve) reproduces this predicted scaling with $T^2$. Temporally coherent emission also occurs at other angles, in the vicinity of the resonant frequencies, as given by Eq.~(\ref{eq:resonant}). The resonant frequencies have a distinct functional dependence with the angle of emission when $v_c>c$. Figure~\ref{fig:undulator}c demonstrates the distinct spectral profile for superluminal undulator radiation. Here, all resonant harmonics asymptotically converge to the Cherenkov angle as $\omega \rightarrow \infty$, exactly as predicted by Eq.~(\ref{eq:resonant}). This feature is unique to superluminal undulator radiation. The spectral intensity at any of the resonant frequencies also grows with $I \propto T^2 \propto N^2$ (inset in Fig.~\ref{fig:undulator}b, yellow). Because the amplitude of the quasiparticle oscillation is large, multiple resonant frequency harmonics result in a broadband (single-cycle) pulse train, seen in the inset of Fig.~\ref{fig:undulator}b. An enhancement in the superluminal undulator radiation occurs when resonant harmonics from multiple quasiparticles cross (see highlighted region in Fig.~\ref{fig:undulator}c). 

\begin{figure}[h!]
\includegraphics[width=1.0\columnwidth]{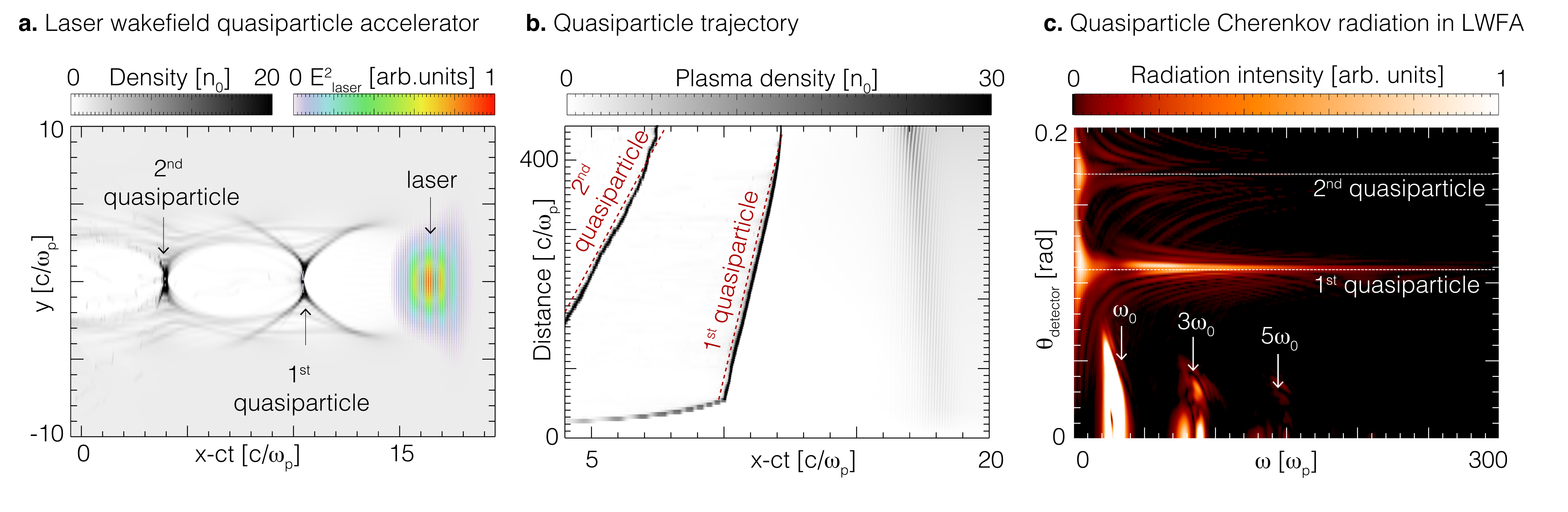}
\caption{\label{fig:lwfa} Quasiparticle Cherenkov radiation in a laser wakefield accelerator. a. Laser driven plasma wakefields. Plasma density appears in gray, and the square of the laser electric field in rainbow colours. b. Waterfall plot showing the quasiparticle trajectories in frame moving at $c$. Each horizontal line corresponds to an on-axis lineout of the plasma density as a function of propagation distance. c. Corresponding spectral intensity in a virtual detector in the far field. The white dashed lines are placed at the Cherenkov angles corresponding to the velocity of the first and second quasiparticles. The fundamental laser frequency, and its high harmonics are also visible.
}
\end{figure}

Figure~\ref{fig:lwfa} shows that coherent quasiparticle emission can also be experimentally realized in laser wakefield acceleration (LWFA), which uses an intense laser pulse to excite a nonlinear wakefield in plasma. Figure~\ref{fig:lwfa}a, which predicts the laser-driven wakefield structure, shows the characteristic electron spikes that play the role of quasiparticles. Compared to the beam-driven case, velocity control can be more challenging in a LWFA  because the group velocity of the laser pulse depends on the plasma density. Nevertheless, Fig.~\ref{fig:lwfa}b demonstrates that the spatially varying plasma density profile provides excellent control over the quasiparticle velocity. As a result, an optical shock appears at the Cherenkov angle in Fig.~\ref{fig:lwfa}c, exactly as predicted. The radiated energy produced by the quasiparticle in Fig.~\ref{fig:lwfa} is three times higher than the energy radiated into the third harmonic of the laser frequency, demonstrating that the process produces clear, experimentally detectable signatures.



Plasma density ramps are essential for experimental demonstration of coherent emission from quasiparticles in plasma accelerators. Plasma density ramps have long been experimentally demonstrated (e.g., in downramp injection in LWFA) by tilting the gas jet with respect to the propagation axis. Corrugated plasmas have also been experimentally demonstrated using two methods~\cite{bib:layer_prl_2007,bib:layer_oe_2009}. The first uses an axicon to map radial intensity modulations of an incident pulse to a line focus with longitudinal intensity modulations. The line-focused pulse preferentially ionizes and heats a clustered gas where its intensity is high, which after, hydrodynamic expansion results in a corrugated plasma channel. The second method uses periodically placed wires distributed along a slot gas jet to obstruct the flow of the clustered gas. The gas density is higher where it is unobstructed, resulting in density modulations upon ionization and heating by a laser pulse. Using these methods, modulations with $35-300~\mathrm{\mu m}$ periods with contrast ratios (modulation depths) in excess of 0.9 have been generated in densities relevant to our work (i.e. $n \simeq 10^{18}~\mathrm{cm}^{-3}$).

The quasiparticle trajectory may also be suitably controlled by using laser pulses with a flying focus~\cite{bib:froula_nphoton_2018,bib:marie_optica_2017} or Arbitrarily Structured Laser Pulses (ASTRL)~\cite{bib:pierce_prr_2023}, where the laser intensity peak can move along an arbitrary trajectory. At low intensities, these pulses may enable quasiparticle-based coherent light sources in nonlinear optics. A generalization of our concept for nonlinear optics is possible by replacing the current density $\mathbf{j}$ by the time derivative of the polarisation density of a nonlinear optical system, i.e., by making the substitution $\mathbf{j} \rightarrow \partial_t\mathbf{P}$, where $\mathbf{P} = \epsilon_0 \chi \mathbf{E}$ and $\chi$ is the susceptibility tensor, which can depend nonlinearly on the electric field $\mathbf{E}$. Similarly, ordered arrangements of spin can be interpreted as quasiparticles and analyzed using Eqs. (\ref{eq:collective}) and (\ref{eq:weight}) by substituting $\mathbf{j} \rightarrow \nabla \times \mathbf{M}$, where $\mathbf{M}$ is the magnetization. 

The quasiparticle concept enables a new class of light sources by bringing temporal coherence and superradiance to configurations where they would otherwise be impossible. Quasiparticles emerge from the collective motion of classical point-like particles, but unlike these particles, quasiparticles have to the flexibility to execute arbitrary trajectories, free from the constraints of classical mechanics, relativity, or existing technology. In the context of plasma-based accelerators, the quasiparticle trajectory can be controlled to produce single-cycle optical shocks with a structured bandwidth that extends from the THz to the extreme ultraviolet. This includes the optical Cherenkov radiation and novel longitudinal undulator regime described here. The flexibility in the motion of quasiparticles can also be used to explore effects such as the reversed Doppler-shift predicted for superluminal particles~\cite{bib:shi_nphysics_2018} or to provide a surrogate for studying radiation from the extreme accelerations experienced in the vicinity of exotic astrophysical objects, such as black holes. Moreover, the quasiparticle concept introduces additional degrees of freedom that can be used to enhance existing single particle radiation mechanisms, such as Smith Purcell, transition, or synchrotron radiation. Essential to all of these is a creative exploration of current density profiles and trajectories. In the near term, quasiparticle radiation can be realized using widely available experimental resources, making it suitable for experimental demonstration at existing laser and plasma accelerator laboratories. 







\acknowledgements

The authors would like to acknowledge very fruitful interactions and discussions with Lu\'is Silva, Robert Bingham, and Raoul Trines. We acknowledge use of the Marenostrum (Spain) and LUMI (Finland) supercomputers through PRACE/EuroHPC awards. Work of BM is supported by Funda\c c\~ao para a Ci\^encia e a Tecnologia (FCT, Portugal) grant number PD/BD/150409/2019. The work of DR, KW, and JPP is supported by the Office of Fusion Energy Sciences under Award Number DE-SC0019135 and DE-SC00215057, the Department of Energy National Nuclear Security Administration under Award Number DE-NA0003856, the University of Rochester, and the New York State Energy Research and Development Authority. The work of JP and WBM is supported by the National Science Foundation award 2108970.

\par

\par 
\nocite{*}


\end{document}